# Inelastic scattering of electrons by metastable hydrogen atoms in a laser field

Gabriela Buica

*Institute of Space Sciences, P.O. Box MG-23,*

*Ro 77125, Bucharest-Măgurele, Romania*

(Dated: September 2, 2015)

## Abstract

The *inelastic* scattering of fast electrons by metastable hydrogen atoms in the presence of a linearly polarized laser field is theoretically studied in the domain of field intensities below $10^{10}$ W/cm$^2$. The interaction of the hydrogen atom with the laser field is described by first-order time-dependent perturbation theory, while the projectile electrons interacting with the laser field are described by the Gordon-Volkov wave functions. An analytic expression is obtained for the differential scattering cross section in the first-order Born approximation for laser-assisted inelastic $e^-$-H($2s$) scattering for the $2s \to nl$ excitation. Detailed analytical and numerical results are presented for inelastic scattering accompanied by one-photon absorption, and the angular dependence and resonance structure of the differential cross sections is discussed for the $2s \to 4l$ excitation of metastable hydrogen.

Keywords:

## I. INTRODUCTION

The scattering of electrons by atoms has been studied from the beginning of the last century starting with the Franck and Hertz experiments and the early theoretical works by Massey and Mohr [1]. During the last few decades the study of electron-atom collisions in the presence of a laser field has been the subject of intense research activities, because of the importance in applied domains such as astrophysics [2], laser and plasma physics [3], or fundamental atomic collision theory, etc. The essence of laser-assisted scattering is that the projectile electron is now allowed to absorb or emit photons during the scattering processes by atoms. In contrast, a free electron alone cannot exchange photons with a laser field unless an atom (third body) is present. Obviously, the laser-assisted scattering processes can provide new aspects regarding the scattering dynamics, the net effect of the laser field being the suppression of the scattering cross section and redistribution of the scattering signal to the $N$-photon scattering channels. Detailed reports on the laser assisted electron-atom collisions processes can be find in the review papers [4–6] and books [7, 8].

Lately, due to advance of new experimental techniques there is a renewed interest in studying the electron-atom scattering in the presence of a laser field for both *elastic* and *inelastic* processes. We stress that for elastic scattering the initial and final atomic states are the same, while for inelastic scattering the final atomic state differs from the initial one. For the inelastic process we should point out two types of collisions when the kinetic energy of the incident electron is above or below the first ionization threshold. For the later case both collisional and radiative interactions can be comparable in strength and simultaneously electron-photon excitation of atom might occur [4, 9]. For weak laser fields, as long as the photon energy is low, the laser-atom interaction might be neglected and the atom can be modeled by a center of force described through a static potential [10, 11]. Obviously, this approximation can be suitable for laser-assisted elastic scattering, where most of the studies are done, in which the atom does not change its state after collision, but for laser-assisted *inelastic* scattering the whole atomic structure must be taken under consideration. It is important to evaluate the contribution of laser-assisted *inelastic* electron-atom scattering because in experimental studies [20–23] it might be quite difficult to separate the signal of elastic and inelastic scattering channels since we have to simultaneous observe the final state of the target, the projectile electron, and the photon. Theoretical works on the laser-assisted inelastic scattering of electrons by the hydrogen atom in its *ground state* by taking into account the atomic "dressing" (i.e. the dipole distortion of the atom by the laser field) in first-order perturbation theory were performed by Jetzke and co-workers [12], Francken and co-workers [13], Bhattacharya and co-workers [14], and Cionga and Florescu [15]. Lately Voitkiv and co-workers [16] have considered laser-assisted inelastic collisions of relativistic electrons with atomic targets in their ground state. An analytical formula was derived for the scattering cross-section using an approach which takes into account the internal degrees of freedom of the target for laser parameters that do not directly influence the atomic target. Very recently, there is an increased interest on the simultaneous electron-photon excitation of helium, in its $^1S$ ground state, in the presence of a laser field [17, 18] by using a nonperturbative R-matrix Floquet theory or a semiperturbative method in the second-order Born approximation [19].

To our knowledge, the inelastic electron-atom scattering process in the presence of a laser field has been theoretically studied in less detail for atoms in *excited states*. It is well known that the atomic dressing effects for elastic scattering processes are proportional to the static dipole polarizability [24–26] and from Radzig and Smirnov's tables [27] the static dipole polarizability of the $nl$ subshell of hydrogen are written as $\alpha_{nl} = n^4 \left(n^2 + 7l^2 + 7l + 14\right)/4$, where $n$ and $l$ are the principal and orbital quantum numbers. Since the static dipole polarizability scales as $n^6$, similar physical effects tend to occur for atoms initially in excited states compared to their ground state but at much lower field intensities [28]. Therefore by increasing the atomic excitation the laser-dressing effects induced by the dipole polarizability should become of larger importance, with a larger probability for experimental verifications. Theoretical calculations on laser-assisted inelastic scattering of electrons by excited hydrogen were performed by Vučić [29], for the $2s \to nl$ and $2p \to nl$ excitations of the $n = 2$ and 3 levels at resonant photon energies, using a Born-Floquet theory. Numerical results on electron-impact $2s \to nl$ ($n = 3$ and 4) excitation of hydrogen in presence of a circularly polarized laser field were published by Purohit and Mathur [30], in the approach of a two-level model atom and rotating wave approximation, for the particular case of a resonant photon energy that matches the $2s - 3p$ optical transition in hydrogen.

The purpose of this paper is to study the $2s \to nl$ inelastic scattering of fast electrons by hydrogen in the metastable $2s$ state in the presence of a linearly polarized laser field. In Sec. II is presented the physical framework used to derive an analytic formula for the differential cross section (DCS) for excitation of an arbitrary state. Since the scattering process under investigation is a three-body problem, i.e., projectile electron, atomic target, and photon, naturally the theoretical treatment presents considerable difficulties and several assumptions are made. (i) Moderate field intensities below $10^{10}$ W/cm$^2$ and fast projectile electrons are considered in order to safely neglect the second-order Born approximation in the scattering potential and the exchange scattering [13]. The interaction between the fast projectile electron and the hydrogen atom is treated in the first-order Born approximation [31]. (ii) The interaction between the projectile electrons and the laser field is described by a Gordon-Volkov wave function [32]. (iii) The dressing of the hydrogen atom by the laser field, i.e., the modification of the target atom in the laser field, is described within the first-order time-dependent perturbation theory (TDPT) in the field [33]. Working in the approximation approach described in [15] and [31] we derive an *analytic formula* for DCS in the laser-assisted inelastic scattering of fast electrons by H(2s) accompanied by one-photon exchange. Section III is devoted to discussion of the numerical results where the angular distributions and the resonance structure of the DCS's are analyzed for the excitation of the $n=4$ subshells. In comparison to the earlier theoretical works [29, 30] the present approach provides a compact analytical formula for DCS that includes the atomic dressing effects beyond the two-level model and does not require the use of the rotating-wave approximation, being applicable for any polarization of the laser field. Atomic units are used throughout this paper unless otherwise specified.

## II. BASIC THEORY AT MODERATE LASER INTENSITIES

The laser-assisted inelastic scattering of electrons by the metastable hydrogen atoms can be formally represented as

$$e^-(E_{k_i}, \mathbf{k}_i) + \text{H}(2s) + N_i\, \gamma\,(\omega, \boldsymbol{\epsilon}) \rightarrow e^-(E_{k_f}, \mathbf{k}_f) + \text{H}(nlm) + N_f\, \gamma\,(\omega, \boldsymbol{\epsilon}), \qquad (1)$$

where $E_{k_i}$ ($E_{k_f}$) and $\mathbf{k}_i$ ($\mathbf{k}_f$) represent the kinetic energy and the momentum vector of the projectile electron in its initial (final) state. H(2s) and H($nlm$) denote the hydrogen atom that is initially in its metastable $2s$ state and finally, after collision, is excited to a state defined by the quantum numbers $n, l$, and $m$, where $m$ is the magnetic quantum number. Here $\gamma$ represents a photon with an energy $\omega$ and a unit polarization vector $\boldsymbol{\epsilon}$, and $N = N_i - N_f$ is the net number of exchanged photons between the projectile electron-atom system and the laser field. The laser field is treated classically and is considered to be a monochromatic electric field

$$\boldsymbol{\mathcal{E}}(t) = \frac{i}{2}\mathcal{E}_0\boldsymbol{\epsilon}\exp(-i\omega t) + cc, \qquad (2)$$

where $\mathcal{E}_0$ represents the peak amplitude of the electric field. In literature, this scattering process (1) is called *free-free transition* since the projectile electron is free both before and after the scattering process. In addition the process is considered *inelastic* since the initial and final states of the target are not identical and the projectile electron energies satisfy the following conservation relation: $E_{k_f} = E_{k_i} + E_{2s} - E_{nlm} + N\omega$, where $E_{2s}$ and $E_{nlm}$ represent the energy of the $2s$ and $nlm$ excited states. Obviously the number of exchanged photons cannot be smaller than a minimal value that is the integer of $N_{min} = (E_{nlm} - E_{2s} - E_{k_i})/\omega$. The kinetic energy spectrum of the scattered electrons consists of a *central line* that corresponds to $N = 0$ (i.e., the total number of absorbed and emitted photons is zero) and a number of *sidebands* with $N = \pm 1, \pm 2, \pm 3,...$, where each pair of *sidebands* corresponds to the scattered electrons with $|N|$ photons absorbed if $N > 0$ (inverse bremsstrahlung) or emitted if $N < 0$ (stimulated bremsstrahlung) [4].

### A. Projectile electron and atomic wave functions

As already mentioned, we consider moderate field intensities and fast projectiles, which imply that the strength of the laser field is smaller than the Coulomb field experienced by the electron in the second Bohr orbit and the velocity of the projectile electron is much larger than the velocity of the bound electron in its second Bohr orbit [5], respectively. The interaction between the laser field and the projectile electron is treated by the Gordon-Volkov wave function [32], and the initial and final states of the scattered electron are described by

$$\chi_{\mathbf{k}}(\mathbf{R}, t) = (2\pi)^{-3/2} \exp\left[-iE_k t + i\mathbf{k} \cdot \mathbf{R} - i\mathbf{k} \cdot \boldsymbol{\alpha}(t)\right], \qquad (3)$$

where $\mathbf{R}$ denotes the position vector of the projectile electron and $\boldsymbol{\alpha}(t)$ describes the classical oscillation motion of the projectile electron in the electric field given by Eq. (2)

$$\boldsymbol{\alpha}(t) = \boldsymbol{\alpha_0} \sin(\omega t), \qquad (4)$$

with the peak amplitude $\boldsymbol{\alpha_0} = \boldsymbol{\epsilon}\sqrt{I}\,\omega^{-2}$, where $I = \mathcal{E}_0^2$ denotes the laser intensity. In Eq. (3) the terms which are proportional to the ponderomotive energy $U_p = I/4\omega^2$ are not included since all calculations presented in the paper are made for moderate intensities below $10^{10}$ W/cm$^2$. For example, at a laser intensity of $10^{10}$ W/cm$^2$ and a photon energy of 1.17 eV the ponderomotive energy is about 0.001 eV and therefore can be safely neglected compared to the projectile energy $E_k$ and photon energy $\omega$.

As long as the field intensity remains moderate the interaction of the laser field with the hydrogen atom is considered within the first-order TDPT. The approximate solution in the first-order TDPT for an electron bound to a Coulomb potential in the presence of an weak external electric field reads

$$\Psi_{nlm}(\mathbf{r},t) = \exp(-iE_n t)\left[\psi_{nlm}(\mathbf{r},t) + \psi_{nlm}^{(1)}(\mathbf{r},t)\right], \qquad (5)$$

where $\mathbf{r}$ represents the position vector of the bound electron, $E_n$ are the Bohr energies, $\psi_{nlm}$ is an unperturbed excited state wave function of the hydrogen, and $\psi_{nlm}^{(1)}$ represents the first-order radiative correction to the atomic wave function. We use the following expression of the first-order correction, as is described in [33]

$$\psi_{nlm}^{(1)}(\mathbf{r},t) = -\frac{\omega}{2}\left[\boldsymbol{\alpha_0}\cdot\mathbf{w}_{nlm}(\Omega_n^+;\mathbf{r})\exp(-i\omega t) + \boldsymbol{\alpha_0}\cdot\mathbf{w}_{nlm}(\Omega_n^-;\mathbf{r})\exp(i\omega t)\right], \qquad (6)$$

with the linear-response vectors $\mathbf{w}_{nlm}$ defined by

$$\mathbf{w}_{nlm}(\Omega_n^\pm;\mathbf{r}) = -G_C(\Omega_n^\pm)\,\mathbf{p}\,\psi_{nlm}(\mathbf{r}), \qquad (7)$$

where $\mathbf{p}$ denotes the momentum operator of the bound electron, $G_C$ is the Coulomb Green's function, and the energies $\Omega_n^\pm$ have the following values

$$\Omega_n^\pm = E_n \pm \omega. \qquad (8)$$

For the hydrogen atom the linear-response vectors $\mathbf{w}_{nlm}$ are expressed as

$$\mathbf{w}_{nlm}(\Omega_n^\pm;\mathbf{r}) = i\left[-\sqrt{\frac{l+1}{2l+1}}\mathcal{B}_{nll+1}(\Omega_n^\pm;\mathbf{r})\mathbf{V}_{l+1lm}(\hat{\mathbf{r}}) + \sqrt{\frac{l}{2l+1}}\mathcal{B}_{nll-1}(\Omega_n^\pm;\mathbf{r})\mathbf{V}_{l-1lm}(\hat{\mathbf{r}})\right], \qquad (9)$$

where the radial functions $\mathcal{B}_{nll'}$, with $l' = l\pm 1$, are evaluated in Ref. [33] using the Coulomb Green's function including both bound and continuum eigenstates. $\mathbf{V}_{l'lm}$ represent the vector spherical harmonics which are defined in Ref. [34] and are equivalent to the $\mathbf{Y}_{ll'm}$ vectors as defined by Edmonds [35], and $\hat{\mathbf{r}} = \mathbf{r}/|\mathbf{r}|$.

### B. Scattering matrix

To proceed further, once we have obtained the atomic and projectile electron wave function in the laser field we are able to derive the scattering matrix for the electron-atom scattering in the static potential

$$V(r, R) = -\frac{1}{R} + \frac{1}{|\mathbf{r} - \mathbf{R}|}. \tag{10}$$

We assume fast projectile electrons such that the scattering process can be well-treated within the first-order Born approximation in the scattering potential $V(r, R)$ and we use a semiperturbative treatment for the scattering process similar to the one developed by Byron and Joachain [31]. Restricting our calculation to the domain of high scattering energies the scattering matrix [13] reads

$$S_{fi} = -i \int_{-\infty}^{\infty} dt \left\langle \chi_{\mathbf{k}_f}(\mathbf{R}, t) \Psi_{n_f l_f m_f}(\mathbf{r}, t) \middle| V(\mathbf{r}, \mathbf{R}) \middle| \chi_{\mathbf{k}_i}(\mathbf{R}, t) \Psi_{n_i l_i m_i}(\mathbf{r}, t) \right\rangle, \tag{11}$$

where $\chi_{\mathbf{k}_i}$ ($\chi_{\mathbf{k}_f}$) represents the Gordon-Volkov wave function for the initial (final) state of the projectile electron in the laser field and $\Psi_{n_i l_i m_i}$ ($\Psi_{n_f l_f m_f}$) is the initial (final) wave function of the bound electron in the laser field, which are given by Eqs. (3) and (5), respectively. For fast projectile electrons the exchange effects are neglected [13] and are not included in Eq. (11). By developing the exponential term of the Gordon-Volkov wave functions $\chi_{\mathbf{k}_i}$ and $\chi_{\mathbf{k}_f}$ in terms of the Bessel functions $J_N$ using the generating function

$$\exp[i\,x\sin(\omega t)] = \sum_{N=-\infty}^{+\infty} J_N(x) \exp(iN\omega t), \tag{12}$$

and substituting Eq. (3) into Eq. (11) we obtain, after integrating over time and projectile electron coordinate, the scattering matrix $S_{fi}$ for the $N$-photon laser-assisted inelastic $e^-$-H($2s$) scattering

$$S_{fi}(N) = -2\pi i \sum_{N=-\infty}^{+\infty} T_{nlm}(N)\,\delta(E_{k_f} + E_n - E_{k_i} - E_{2s} - N\omega), \tag{13}$$

where $\delta$ is the Dirac delta function that assures the energy conservation. Here $T_{nlm}(N)$ represents the total transition amplitude for a $N$-photon inelastic scattering process, which can be written as a sum of two terms:

$$T_{nlm}(N) = T_{nlm}^{(0)}(N) + T_{nlm}^{(1)}(N). \tag{14}$$

In particular, for $n = 2$, $l = 0$, and $m = 0$ the scattering matrix reduces to the one of the laser-assisted elastic scattering process [26]. The first term, $T_{nlm}^{(0)}$, in the right-hand side of Eq. (14) denotes the projectile electron transition amplitude and is related to the Bunkin-Fedorov formula [10] in which the atomic dressing is neglected, $T_{nlm}^{(1)}(N) \simeq 0$,

$$T_{nlm}^{(0)}(N) = J_N(\boldsymbol{\alpha_0} \cdot \mathbf{q})\langle\psi_{nlm}|F(\mathbf{q})|\psi_{2s}\rangle, \tag{15}$$

where $F(\mathbf{q})$ is the generalized form factor

$$F(\mathbf{q}) = \frac{1}{2\pi^2 q^2}\left[\exp(i\mathbf{q}\cdot\mathbf{r}) - \delta_{n2}\delta_{l0}\delta_{m0}\right], \tag{16}$$

and $\mathbf{q}$ represents the momentum transfer vector of the projectile electron, i.e., $\mathbf{q} = \mathbf{k}_i - \mathbf{k}_f$. The momentum transfer $q = |\mathbf{q}|$ varies between the following boundaries $k_i - k_f \leq q \leq k_i + k_f$ for forward and backward scattering. We should note that in Eqs. (15) the field and the projectile electron contributions to transition amplitude $T_{nlm}^{(0)}$ are completely decoupled, since the laser field dependence of the electronic transition amplitude is contained in the argument of the Bessel function, $\boldsymbol{\alpha_0} \cdot \mathbf{q}$, only. $T_{nlm}^{(0)}$ describes the direct excitation of H(2s) by projectile electron interaction. We remind that the Bunkin-Fedorov formula is calculated within the first-order Born approximation, while the well known Kroll-Watson formula [11] is derived beyond the first-order Born approximation with the projectile electron momentum and energy shifted. Both Bunkin-Fedorov and Kroll-Watson approximations are are unappropriated for laser-assisted inelastic scattering since are derived for laser-assisted scattering of an electron by a static potential.

The last term, $T_{nlm}^{(1)}$, in the right-hand side of the inelastic transition amplitude Eq. (14), denotes the atomic transition amplitude and occurs due to modification of the atomic state by the laser field described by the first-order radiative corrections $\psi_{nlm}^{(1)}(\mathbf{r},t)$. One of the $N$ photons is exchanged (it may be emitted or absorbed) between the laser field and the bound electron. The general structure of $T_{nlm}^{(1)}$ is written as

$$T_{nlm}^{(1)}(N) = -\frac{\alpha_0\omega}{2}\left[J_{N-1}(\boldsymbol{\alpha_0}\cdot\mathbf{q})\mathcal{M}_{at}^{(1)}(\Omega_2^+,\Omega_n^-,\mathbf{q}) + J_{N+1}(\boldsymbol{\alpha_0}\cdot\mathbf{q})\mathcal{M}_{at}^{(1)}(\Omega_2^-,\Omega_n^+,\mathbf{q})\right], \tag{17}$$

where $\mathcal{M}_{at}^{(1)}(\Omega_2^+,\Omega_n^-,\mathbf{q})$ and $\mathcal{M}_{at}^{(1)}(\Omega_2^-,\Omega_n^+,\mathbf{q})$ denote the following first-order atomic transition amplitudes involving initial and final atomic states only

$$\mathcal{M}_{at}^{(1)}\left(\Omega_2^+,\Omega_n^-,\mathbf{q}\right) = \langle\psi_{nlm}|F(\mathbf{q})|\boldsymbol{\epsilon}\cdot\mathbf{w}_{200}(\Omega_2^+)\rangle + \langle\boldsymbol{\epsilon}\cdot\mathbf{w}_{nlm}(\Omega_n^-)|F(\mathbf{q})|\psi_{200}\rangle, \tag{18}$$

and
$$\mathcal{M}_{at}^{(1)}\left(\Omega_2^-, \Omega_n^+, \mathbf{q}\right) = \langle\psi_{nlm}|F(\mathbf{q})|\boldsymbol{\epsilon}\cdot\mathbf{w}_{200}(\Omega_2^-)\rangle + \langle\boldsymbol{\epsilon}\cdot\mathbf{w}_{nlm}(\Omega_n^+)|F(\mathbf{q})|\psi_{200}\rangle. \qquad (19)$$

The first term in the right-hand side of Eqs. (18) and (19) describes the atom interacting first with the laser field followed by the interaction with the projectile electron, while in the second term the projectile electron-atom interaction precedes the atom-laser interaction. For notational simplicity the **r**-dependency is not shown in the above equations. We recall that the vectors $\mathbf{w}_{nlm}$ are given by Eq. (7) and the energies $\Omega_n^+$ ($\Omega_n^-$) corresponding to one-photon absorption (emission) are defined in Eq. (8).

### C. One-photon scattering process

Now, we focus our investigations on the scattering process described by Eq. (1) in which one photon is exchanged by the colliding system and we present specific formulas for one-photon absorption ($N = 1$) and emission ($N = -1$) processes. Strictly speaking, however, we note that the calculations of the $|N| > 1$ processes require that the laser-atom interaction should be treated at least to the second-order in the field [36, 37], unless we are interested in DCS's at large scattering angles where the atomic dressing is negligible. Since our formulas are derived up to the first-order only for the atomic dressing, we consider one-photon ($N = \pm1$) processes from now on. Whenever the argument of the Bessel functions is small, i.e. $\boldsymbol{\alpha_0}\cdot\mathbf{q} \ll 1$, which is satisfied at small scattering angles with moderate laser intensities or at any scattering angles with low laser intensities, the following approximate relation for the Bessel functions $J_N(\boldsymbol{\alpha_0}\cdot\mathbf{q}) \simeq (\boldsymbol{\alpha_0}\cdot\mathbf{q}/2)^N N!^{-1}$ may be used. In addition, we must only keep the leading terms in the calculation of the electronic and atomic transition amplitudes Eqs. (15) and (17). Substituting the partial wave expansion of the exponential term, $\exp(i\mathbf{q}\cdot\mathbf{r})$, in the form factor Eq. (15) after performing the angular integration, the electronic transition amplitude $T_{nlm}^{(0)}$ reads

$$T_{nlm}^{(0)}(N=1) = \frac{1}{(2\pi)^2}\frac{\boldsymbol{\alpha_0}\cdot\mathbf{q}}{2}f_{el}^{B1}(q), \qquad (20)$$

where $f_{el}^{B1}(q)$ is the first-order Born approximation of the scattering amplitude for field-free inelastic electron scattering by hydrogen. Its evaluation for $2s \to nlm$ excitation yields

$$f_{el}^{B1}(q) = -\frac{4\sqrt{\pi}\,i^l}{q^2}Y_{lm}^*(\hat{\mathbf{q}})\mathcal{I}_{nl}(q), \qquad (21)$$

where $Y_{lm}$ are the spherical harmonics, $\hat{\mathbf{q}} = \mathbf{q}/|\mathbf{q}|$, and $\mathcal{I}_{nl}$ represents an electronic radial integral defined by

$$\mathcal{I}_{nl}(q) = \int_0^\infty dr \ r^2 R_{nl}(r) j_l(qr) R_{20}(r) - \delta_{n2}\delta_{l0}\delta_{m0}, \tag{22}$$

where $j_l(qr)$ represents the spherical Bessel functions and $R_{nl}(r)$ is the hydrogenic radial function. An analytic expression for the first term in the right-hand side of the electronic radial integral $\mathcal{I}_{nl}$ is given by Eq. (A2) of Appendix A.

Next, by using the partial wave expansion of the exponential $\exp{(i\mathbf{q}\cdot\mathbf{r})}$ and the vector $\mathbf{w}_{nlm}$ definition given by Eq. (9), after performing the angular integration, we obtain for the first term in the right-hand side of the first-order atomic transition amplitude Eq. (18),

$$\langle \psi_{nlm} | \frac{\exp{(i\mathbf{q}\cdot\mathbf{r})}}{2\pi^2 q^2} | \boldsymbol{\epsilon} \cdot \mathbf{w}_{200}(\Omega_2^+) \rangle = \frac{i^l}{\pi^{3/2} q^2} \left[ \sqrt{\frac{l}{2l+1}} \mathcal{T}_{nlm}^{l-1,a}(\Omega_2^+, q) + \sqrt{\frac{l+1}{2l+1}} \mathcal{T}_{nlm}^{l+1,a}(\Omega_2^+, q) \right], \tag{23}$$

where

$$\mathcal{T}_{nlm}^{l',a}(\Omega_2^+, q) = \boldsymbol{\epsilon} \cdot \mathbf{V}_{l'lm}^*(\hat{\mathbf{q}}) \ \mathcal{J}_{nll',20}^a(\Omega_2^+, q), \tag{24}$$

and $\mathcal{J}_{nll',20}^a$ is an atomic radial integral defined as

$$\mathcal{J}_{nll',20}^a(\Omega, q) = \int_0^\infty dr \ r^2 R_{nl}(r) j_{l'}(qr) \mathcal{B}_{201}(\Omega; r). \tag{25}$$

Similarly, for the second term in the right-hand side of Eq. (18) we obtain

$$\langle \boldsymbol{\epsilon} \cdot \mathbf{w}_{nlm}(\Omega_n^-) | \frac{\exp{(i\mathbf{q}\cdot\mathbf{r})}}{2\pi^2 q^2} | \psi_{200} \rangle = -\frac{i^l}{\pi^{3/2} q^2} \left[ \sqrt{\frac{l}{2l+1}} \mathcal{T}_{nlm}^{l-1,b}(\Omega_n^-, q) + \sqrt{\frac{l+1}{2l+1}} \mathcal{T}_{nlm}^{l+1,b}(\Omega_n^-, q) \right], \tag{26}$$

where

$$\mathcal{T}_{nlm}^{l',b}(\Omega_n^-, q) = \boldsymbol{\epsilon} \cdot \mathbf{V}_{l'lm}^*(\hat{\mathbf{q}}) \ \mathcal{J}_{nll',20}^b(\Omega_n^-, q), \tag{27}$$

and $\mathcal{J}_{nll',20}^b$ is another atomic radial integral that is defined as

$$\mathcal{J}_{nll',20}^b(\Omega, q) = \int_0^\infty dr \ r^2 R_{20}(r) j_{l'}(qr) \mathcal{B}_{nll'}(\Omega; r). \tag{28}$$

The analytic expressions of the radial integrals $\mathcal{J}_{nll'}^a$ and $\mathcal{J}_{nll'}^b$, with $l' = l \pm 1$ (if $l > 0$) and $l' = 1$ (if $l = 0$), are given by Eqs. (B4) and (B6) of Appendix B. The atomic transition amplitude $T_{nlm}^{(1)}$ is obtained from Eq. (17) by substituting Eqs. (23) and (26) into Eqs. (18) and (19) as

$$T_{nlm}^{(1)}(N=1) = -\frac{i^l \alpha_0 \ \omega}{2\pi^{3/2} q^2} \left[ \sqrt{\frac{l}{2l+1}} \mathcal{T}_{nlm}^{l-1}(\Omega_2^+, \Omega_n^-, q) + \sqrt{\frac{l+1}{2l+1}} \mathcal{T}_{nlm}^{l+1}(\Omega_2^+, \Omega_n^-, q) \right], \tag{29}$$

for one-photon absorption ($N = 1$), where we have introduced the following notation

$$\mathcal{T}^{l'}_{nlm}(\Omega_2^{\pm}, \Omega_n^{\mp}, q) = \boldsymbol{\epsilon} \cdot \mathbf{V}^*_{l'lm}(\hat{\mathbf{q}}) \mathcal{J}_{nll'}(\Omega_2^{\pm}, \Omega_n^{\mp}, q). \tag{30}$$

The atomic radial integral $\mathcal{J}_{nll'}$ is defined as the difference of the two atomic radial integrals Eqs. (25) and (28)

$$\mathcal{J}_{nll'}(\Omega_2^{\pm}, \Omega_n^{\mp}, q) = \mathcal{J}^a_{nll',20}(\Omega_2^{\pm}, q) - \mathcal{J}^b_{nll',20}(\Omega_n^{\mp}, q), \tag{31}$$

with $l' = l \pm 1$ (if $l > 0$) and $l' = 1$ (if $l = 0$).

Similarly, for one-photon emission ($N = -1$) the atomic transition amplitude is derived as

$$T^{(1)}_{nlm}(N = -1) = -\frac{i^l \alpha_0 \, \omega}{2\pi^{3/2} q^2} \left[ \sqrt{\frac{l}{2l+1}} \, \mathcal{T}^{l-1}_{nlm}(\Omega_2^-, \Omega_n^+, q) + \sqrt{\frac{l+1}{2l+1}} \, \mathcal{T}^{l+1}_{nlm}(\Omega_2^-, \Omega_n^+, q) \right]. \tag{32}$$

It is clear from Eqs. (29) and (32) that the one-photon atomic transition amplitudes involve intermediate states with angular momentum $l' = l \pm 1$, where $l$ is the angular momentum of the final state.

### D. Differential cross section for the excitation of the $nl$ subshell

Finally, the differential cross section for the laser-assisted inelastic $e^- + \mathrm{H}(2s) \to e^- + \mathrm{H}(nl)$ scattering in which the energy of the projectile electron is modified by $E_{2s} - E_n + N\omega$, summed over the magnetic quantum number, $m$, of the final state reads

$$\frac{d\sigma_{nl}(N)}{d\Omega} = (2\pi)^4 \frac{k_f(N)}{k_i} \sum_{m=-l}^{l} |T_{nlm}(N)|^2, \tag{33}$$

where the inelastic transition amplitude $T_{nlm}$ is given by Eq. (14). Within the framework described in the previous subsections the DCS for the inelastic scattering process accompanied by one-photon absorption ($N = 1$) or emission ($N = -1$), summed over the magnetic quantum number of the final state, takes a compact analytic form after some algebra

$$\frac{d\sigma_{nl}(N = \pm 1)}{d\Omega} = \frac{k_f(N = \pm 1)}{k_i} \frac{I}{2 \, \omega^2 \, q^4} \left[ \mathcal{C}_{nl}(\Omega_2^{\pm}, \Omega_n^{\mp}, q) + \frac{|\boldsymbol{\epsilon} \cdot \mathbf{q}|^2}{q^2} \mathcal{D}_{nl}(\Omega_2^{\pm}, \Omega_n^{\mp}, q) \right], \tag{34}$$

where the quantities $\mathcal{C}_{nl}$ and $\mathcal{D}_{nl}$ are defined as follows

$$\mathcal{C}_{nl} = \frac{l(l+1)}{2l+1} (\mathcal{J}_{nll+1} + \mathcal{J}_{nll-1})^2, \tag{35}$$

and

$$\begin{aligned}\mathcal{D}_{nl} &= \frac{(l+1)(l+2)}{2l+1}\mathcal{J}_{nll+1}^2 + \frac{l(l-1)}{2l+1}\mathcal{J}_{nll-1}^2 - \frac{6l(l+1)}{2l+1}\mathcal{J}_{nll+1}\mathcal{J}_{nll-1} \\ &+ \frac{4q}{\omega}\mathcal{I}_{nl}\left[\frac{q}{2\omega}(2l+1)\,\mathcal{I}_{nl} + (l+1)\mathcal{J}_{nll+1} - l\mathcal{J}_{nll-1}\right]. \end{aligned} \quad (36)$$

For notational simplicity, in the above formulas we drop off the arguments $\Omega$ and $q$ of the quantities $\mathcal{C}_{nl}$ and $\mathcal{D}_{nl}$. In Eq. (34) the summation over the magnetic quantum number is performed by taking into account the specific summations relations for the vector spherical harmonics $\mathbf{V}_{l'lm}$ [34] which are presented in the Appendix C. The definitions of quantities $\mathcal{C}_{nl}$ and $\mathcal{D}_{nl}$ given by Eqs. (35) and (36) have a similar form compared to the definitions of $\mathcal{P}$ and $\mathcal{Q}$ reported in Ref. [15] for laser-assisted inelastic $e^-$-H(1s) scattering, but obviously the expressions of electronic $\mathcal{I}_{nl}$ and atomic $\mathcal{J}_{nll'}$ radial integrals are different, since in the present work the hydrogen atom is initially in its metastable $2s$ state. Next, after the derivation of a general formula for the one-photon differential scattering cross section we present some particular cases where Eq. (34) can take quite simple analytic expressions.

1. *Differential cross section for the ns-subshell excitation*

For the inelastic scattering where the final state is a $ns$-subshell ($2s \to ns$ excitation), which implies that both initial and final atomic states have spherical symmetry, we obtain after setting the orbital quantum number $l=0$ in Eq. (34) the following simple formula for DCS:

$$\frac{d\sigma_{n0}(N=\pm 1)}{d\Omega} = I\,\frac{k_f(N=\pm 1)}{k_i}\frac{|\boldsymbol{\epsilon}\cdot\mathbf{q}|^2}{\omega^4 q^4}\left(\mathcal{I}_{n0} + \frac{\omega}{q}\mathcal{J}_{n01}\right)^2. \quad (37)$$

In particular, for $n=2$, the DCS given by Eq. (37) is in perfect agreement with our for the laser-assisted elastic $e^-$-H(2s) scattering [26].

2. *Differential cross section for the np-subshell excitation*

For the inelastic scattering where the final state is a $np$-subshell ($2s \to np$ excitation), which implies that only the initial atomic state has a spherical symmetry, we obtain after setting $l=1$ in Eq. (34) the following formula for DCS:

$$\frac{d\sigma_{n1}(N=\pm 1)}{d\Omega} = I\,\frac{k_f(N=\pm 1)}{k_i}\frac{1}{2\,\omega^2\,q^4}\left\{\frac{2}{3}(\mathcal{J}_{n12}+\mathcal{J}_{n10})^2\right.$$

$$+ \frac{|\boldsymbol{\epsilon} \cdot \mathbf{q}|^2}{q^2} \left[ \frac{6}{5} \mathcal{J}_{n12}^2 - 4\mathcal{J}_{n12}\mathcal{J}_{n10} + \frac{4q}{\omega} \mathcal{I}_{n1} \left( \frac{3q}{2\omega} \mathcal{I}_{n1} + 2\mathcal{J}_{n12} - \mathcal{J}_{n10} \right) \right] \right\} \quad (38)$$

It is clear from Eqs. (37) and (38) that DCS depends on the projectile electron momenta, photon energy, and has a maxim value in the scattering geometry in which the polarization vector is parallel to the momentum transfer, $\boldsymbol{\epsilon} \parallel \mathbf{q}$.

### E. Differential cross section in the limit of negligible field-atom interaction

In the domain of negligible laser-atom interaction ($T_{nlm}^{(1)} \simeq 0$), i.e. at large scattering angles and photon energies that are far away from any (intermediate) resonance, only the projectile electron-laser interaction has to be considered. Again, we employ the partial wave expansion of the exponential term of the form factor $F(\mathbf{q})$, then we substitute Eq. (15) into Eq. (33), and after performing the summation over the final state magnetic quantum number we obtain a simple result for the DCS in the soft-photon limit

$$\frac{d\sigma_{nl}(N)}{d\Omega} = 4 \frac{k_f(N)}{k_i} \frac{2l+1}{q^4} |J_N(\boldsymbol{\alpha_0} \cdot \mathbf{q})|^2 \ \mathcal{I}_{nl}^2, \quad (39)$$

that is the equivalent of the Beigman and Chichkov formula [38] for the laser-assisted excitation of H($2s$) atoms by fast electrons.

In particular, for one-photon exchange, if the dressing of the target is neglected in Eq. (34), i.e. the atomic integrals are neglected ($\mathcal{J}_{nll\pm 1} \simeq 0$) in Eqs. (35) and (36), the following analytic result is derived for the inelastic scattering process

$$\frac{d\sigma_{nl}(N=\pm 1)}{d\Omega} = I \frac{k_f(N=\pm 1)}{k_i} \frac{2l+1}{\omega^4 \ q^4} |\boldsymbol{\epsilon} \cdot \mathbf{q}|^2 \ \mathcal{I}_{nl}^2. \quad (40)$$

Clearly, because the DCS is proportional to $\omega^{-4}$, a factor which arises from the projectile electron quiver amplitude as can be seen in Eq. (4), the infrared $CO_2$ laser (photon energy $\omega = 0.117$ eV) was preferred to Nd:YAG ($\omega = 1.17$ eV) or He:Ne ($\omega = 2$ eV) lasers in the earlier laser-assisted electron-atom experiments. Only recently, laser-assisted experiments using femtosecond near-infrared Ti:sapphire ($\omega = 1.55$ eV) [39] or Nd:YAG lasers [18, 40] were reported.

### III. NUMERICAL EXAMPLES AND DISCUSSIONS

To start with, we check the reliability of our results for the particular case of *elastic* scattering of fast electrons by the hydrogen atoms, in its metastable $2s$ state, in the presence of a linearly polarized laser field of moderate power [26]. We recall that our analytic results are also valid for the case of free-free *elastic* scattering corresponding to the energy conservation $E_{k_f} = E_{k_i} + N\omega$, a process in which the atomic hydrogen remains in its initial state $2s$ and the kinetic energy of the projectile electron changes by an integer multiple of the photon energy $\omega$. We consider a scattering geometry depicted in Fig. 1, denoted as G1, where the laser field is linearly polarized in the same direction along the momentum vector of the ingoing electron $\boldsymbol{\epsilon} \parallel \mathbf{k}_i$, and $\boldsymbol{\epsilon}$ defines the direction of z-axis. For this scattering geometry the argument of the Bessel function can be expressed as $\boldsymbol{\alpha_0} \cdot \mathbf{q} = \sqrt{I}\,\omega^{-2}(k_i - k_f \cos\theta)$, where $\theta$ is the scattering angle between the initial and final momentum vectors of the projectile electron $\mathbf{k}_i$ and $\mathbf{k}_f$. The numerical results for one-photon absorption ($N = 1$) are shown in Figs. 2(a) and 2(b) in terms of the DCS's at the incident projectile electron energy $E_{k_i} = 100$ eV, with the photon energies of 1.17 eV (Nd:YAG laser) and 2 eV (He:Ne laser), respectively. The solid and dashed lines represent the numerical results which include the atomic dressing obtained from Eq. (37) and those given by Eq. (40) where the dressing of the atom is neglected, respectively. The dot-dashed lines represent the atomic dressing contribution to DCS calculated within the first-order TDPT as $(2\pi)^4 k_f/k_i \sum_{m=-l}^{l} |T_{nlm}^{(1)}(N)|^2$. The presented results for the elastic scattering process agree with the previous data shown in Figs. (1) and (2) of Ref. [26]. The first minimum appearing in Fig. 2(a) at the scattering angle of $\theta \simeq 3.8°$ originates from the destructive dynamic interference of the electronic and atomic contributions in Eq. (37). The value of $\theta \simeq 3.8°$ is nothing but the solution of the following equation: $\mathcal{I}_{n0} + \omega\,q^{-1}\mathcal{J}_{n01} = 0$, where the $\theta$-dependency is included in the momentum transfer $q$ and the radial integrals $\mathcal{I}_{n0}$ and $\mathcal{J}_{n01}$. This is the so-called dynamic minimum, whose position depends on the choice of the polarization potential for the target in the laser field. Another type of minimum at $\theta \simeq 6.2°$ and $8.1°$ in Figs. 2(a) and 2(b), respectively, occurs because the scalar product $\boldsymbol{\epsilon} \cdot \mathbf{q}$ vanishes in Eqs. (37) and (40), at the scattering angle given by the relation of $\theta = \arccos(k_i/k_f)$ in the scattering geometry G1. This is the so-called kinematic minimum whose position depends on the polarization geometry and the initial and final energies of the projectile electron, and exists for the photoabsorption ($N > 0$) scattering processes only, where the projectile electron momenta must satisfy the relation of $k_f > k_i$. We have also compared our numerical results, not shown here, for laser-assisted elastic $e^-$-H($2s$) scattering with those published by Vučić and Hewitt [28] based on a Born-Floquet theory and we found out a good quantitative agreement.

Second, we want to check the dynamic atomic polarizability of the $2s$ state of hydrogen, $\alpha_{2s}(\omega)$, since the atomic dressing effect increases with increasing excitation of the atomic target, fact that is reflected in the increasing static polarizability of the excited states [26]. The dynamic polarizability is defined as a function of the photon energy [23, 41], and can be expressed in the limit of small momentum transfer (Bethe-Born approximation) [42], in terms of linear-response vector $\mathbf{w}_{200}$ as

$$\alpha_{2s}(\omega) = -\frac{1}{\omega\,\boldsymbol{\epsilon}\cdot\mathbf{q}} \lim_{q\to 0} \left[ \langle \psi_{200} | \exp(i\mathbf{q}\cdot\mathbf{r}) | \boldsymbol{\epsilon}\cdot\mathbf{w}_{200}(\Omega_2^+)\rangle + \langle \boldsymbol{\epsilon}\cdot\mathbf{w}_{200}(\Omega_n^-)| \exp(i\mathbf{q}\cdot\mathbf{r})|\psi_{200}\rangle \right]. \tag{41}$$

After performing some algebra and recalling Eqs. (23) and (26) the dynamic polarizability of the $2s$ state reads

$$\alpha_{2s}(\omega) = \frac{1}{\omega\,q} \lim_{q\to 0} \mathcal{J}_{201}(\Omega_2^+, \Omega_2^-, q), \tag{42}$$

where the radial integral $\mathcal{J}_{201}(\Omega_2^+, \Omega_2^-, q)$ is defined by Eq. (31). In particular, in the soft-photon limit ($\omega \ll 1$ a.u.) the dynamic polarizability gives the well known value of the static dipole polarizability of the $2s$ state $\alpha_{2s}=120$ a.u.. In Fig. 3 is plotted the dynamic polarizability of the $2s$ state of hydrogen, $\alpha_{2s}(\omega)$, calculated in the limit of small momentum transfer of Eq. (42), as a function of the photon energy. The numerical results are in good agreement with the data presented by Tang and Chan [43] and their references therein. The behavior of the dynamic polarizability of the $2s$ state is qualitatively similar to the one of the ground state [44], and $\alpha_{2s}(\omega)$ changes its sign whenever the photon energy passes through an atomic resonance as well.

Next, we apply the analytic formulas derived in the above sections to calculate the DCS's for the *inelastic* electron scattering by a hydrogen atom in its metastable $2s$ state in the presence of a linearly polarized laser field, and we focus our numerical examples on one-photon absorption ($N = 1$). We show our numerical results for a higher scattering energy $E_{k_i} = 500$ eV and a photon energy corresponding to the Nd:YAG laser $\omega = 1.17$ eV, where we expect important dressing effects for H($2s$). We choose low photon energies and high energies of the projectile electron, such that neither the photon nor the projectile electron can separately excite the atomic upper state. Figures 4(a)-4(d) present the DCS's normalized to the laser intensity for the excitation of the $n = 4$ subshells ($4s, 4p, 4d$, and $4f$) in the scattering geometry G1. The solid and dashed lines correspond to DCS's which include the atomic dressing calculated with Eq. (34) and neglect the atomic dressing given by Eq.

(40), respectively. The dot-dashed lines represent the atomic contribution to DCS given by $(2\pi)^4 k_f/k_i \sum_{m=-l}^{l} |T_{nlm}^{(1)}(N)|^2$. In Fig. 4(a) is also plotted by dot-dot-dashed line the total DCS for transition to any $n = 4$ subshell calculated as $\sum_{l=0}^{n-1} d\sigma_{nl}(N)/d\Omega$. As in the case of elastic scattering the atomic dressing effects are dominant in the forward scattering ($\theta = 0°$) and our numerical results in Figs. 4(a)-4(d) indicate that at small scattering angles ($\theta < 12°$) the atomic contribution to the total DCS is more important than the electronic contribution. The atomic dressing effects are dominant for scattering angles that are $\theta < 4°$ in Fig. 4(a), $\theta < 12°$ in Fig. 4(b), $\theta < 20°$ in Fig. 4(c), and $\theta < 40°$ in Fig. 4(d). Dynamic minima occur for the $2s \to 4s$ excitation only at the following scattering angles $\theta \simeq 3.1°, 7.2°$ and $10.7°$ in Fig. 4(a). As mentioned before, these dynamic minima are independent of the scattering geometry and represent the results of cancellation of the electronic and atomic radial integrals in Eq. (37). Compared with the elastic process we notice the different character of interferences between the electronic and atomic radial integrals in DCS. At photon energy $\omega = 1.17$ eV the cinematic minima are not allowed for any of the $2s \to 4l$ ($l = 0, 1, 2,$ and 3) excitations, because the projectile electron momenta satisfy the relation $k_f < k_i$ for photon energies lower than 2.55 eV and therefore the scalar product $\boldsymbol{\epsilon} \cdot \mathbf{q}$ does not vanish. At larger scattering angles the elastic DCS is not decreasing compared to inelastic DCS due to the nonzero electronic transition amplitude of the electron projectile-nucleus interaction in Eq. (21). Our numerical results based on Eq. (34), not shown here, are in qualitative agreement with the earlier results of DCS's calculated in a two-level approximation by Purohit and Mathur [30] for a circularly polarized laser field, at the resonant photon energy of 1.89 eV which corresponds to the $2s - 3s$ transition in hydrogen.

Figures 5 and 6 show the DCS's with respect of the photon energy for one-photon absorption ($N = 1$) corresponding to the $2s \to 4s$ (short dashed line), $2s \to 4p$ (dot-dashed line), $2s \to 4d$ (long dashed line), and $2s \to 4f$ (solid line) excitations of hydrogen in the scattering geometry G1. The DCS's are normalized to the laser intensity and calculated at the incident projectile electron energy $E_{k_i} = 500$ eV for the scattering angles $\theta = 5°$ in Fig. 5 and $\theta = 30°$ in Fig. 6. At small scattering angle $\theta = 5°$ the atomic dressing is important and the DCS for $4f$ subshell is dominant, while at $\theta = 30°$ the DCS's for $4s$ and $4p$ subshells are dominant, but at non-resonant photon energies the atomic dressing is negligibly small compared to the electronic contribution. The DCS's for $2s \to 4l$ ($l = 0, 1, 2,$ and 3) excitations as a function of the photon energy show a strongly dependence on the atomic structure and exhibits a series of resonances. The first resonance in both Figs. 5 and 6 occurs at the photon energy $\omega \simeq 0.66$ eV that matches the energy difference between the $4s$ and $3p$ states. The origin of this resonance resides in the dipole coupling of a final $4l$ state and a lower intermediate $3l'$ state ($l' = l \pm 1$), as can be noticed from Eq. (26), and is related to the poles of the atomic transition amplitude due to atomic integral $\mathcal{J}^b_{4ll',20}(\Omega_4^-, q)$, given by Eq. (B6), at photon energies such that $\omega = (1/n'^2 - 1/4^2)/2$, with $2 \leq n' < 4$. Another resonance with a similar origin occurs at $\omega \simeq 2.55$ eV due to the dipole coupling of a final $4l$ state and a lower intermediate $2l'$ state ($l' = l \pm 1$). In contrast, the rest of the resonances that appear in Figs. 5 and 6 at photon energies $\omega \simeq 1.88, 2.55, 2.85$ eV, ..., are associated to the $2s \to n'p$ atomic transitions, with $n' > 2$. These resonances occur due to one-photon absorption from the initial $2s$ state to an intermediate $n'p$ state followed by a transition to the final $4l$ state by projectile electron interaction, as can be noticed from Eq. (23). Exactly at resonance the atomic integral $\mathcal{J}^a_{4ll',20}(\Omega_2^+, q)$, given by Eq. (B4), presents poles that occur at photon energies that match atomic resonances such that $\omega = (1/2^2 - 1/n'^2)/2$, with $n' > 2$. It is clear that these two types of resonances occur due to simultaneously projectile electron-photon excitation process of hydrogen. The hydrogen atom is excited to the $4l$ state either (i) by absorbing first some of the projectile electron kinetic energy, then absorbing (or emitting) a photon from (to) the laser field or (ii) by absorbing first a photon from the laser field, then absorbing (or transferring) the energy difference from (to) projectile electron.

Despite the fact there are very new experiments developed on laser-assisted electron-atom scattering [39, 40] these kind of investigations are still quite difficult to perform, especially those involving excited states of atoms [18]. From the experimental point of view [17], we believe that some of these resonances are particularly interesting. Indeed, the laser-assisted excitation processes in $e^-$-H($2s$) scattering accompanied by one-photon absorption may be easier to detect at photon energy close to the 0.66 eV resonance because it does not correspond to a resonance of the *elastic* scattering processes of H($2s$) with one-photon absorption [26].

## IV. SUMMARY AND CONCLUSIONS

We theoretically studied the *inelastic* scattering of fast electrons by H($2s$) in a laser field of moderate intensities, a process that is accompanied by the change of the initial state of the atom. We included the laser dressing of the excited states by using TDPT in the first order of the field and the scattered electron embedded in the laser field is described by the well-known Gordon-Volkov solution. Since we chose high electron energies, the scattering process is treated within the framework of the first-order Born approximation. This formalism includes the dressing effects in laser-assisted inelastic electron-atom scattering beyond the two-level approximation and does not require the use of the rotating-wave approximation, being applicable for any polarization of the laser field. A new *analytic formula* was derived for DCS in the laser-assisted inelastic electron-hydrogen scattering for the $2s \to nl$ excitation of an arbitrary $nl$ state. We compared our results for laser-assisted elastic electron scattering by the metastable $2s$ state of hydrogen with similar calculations [26, 28] and we found a very good agreement, for one-photon absorption. A very good agreement was found out for the dynamic dipole polarizability of the $2s$ state [43]. We found out a qualitative agreement with the inelastic earlier results calculated in a two-level approximation at resonance photon energy [30] showing the accuracy and efficiency of our results. The dynamics of the laser-assisted inelastic $e^-$-H($2s$) collisions accompanied by one-photon absorption differs both qualitatively and quantitatively from that involving the ground state [15] due to the high polarizabilities of the excited states. The presented numerical data for each subshell of the $n = 4$ level indicates that at small scattering angles the atomic dressing effects for inelastic collisions are significantly stronger than for elastic $e^-$-H($2s$) scattering. We found out a significant increase of the DCS's for the optically forbidden transitions $s - s$, $s - d$ and $s - f$ due to simultaneously electron-photon excitation of H($2s$). We discussed the origin of the peaks in the resonance structure of the DCS's as a function of the photon energy for laser-assisted inelastic $e^-$-H($2s$) scattering. For the studied range of photon energies our analysis showed important differences from the dynamics of laser-assisted elastic scattering that are mainly due to the dipole coupling of the final $4l$ state with intermediate $2l'$ and $3l'$ states.

**Acknowledgments**

The work by G. Buica was supported by research programs Laplas 3 PN 09 39N and contract Capacitati-Scafi F03/2014 from the Ministry of Education and Research of Romania.

**Appendix A: Electronic radial integral $\mathcal{I}'_{nl}$**

The electronic radial integral, $\mathcal{I}'_{nl}$, required to calculate the inelastic electronic transition amplitude, $T^{(0)}_{nlm}$, given in Eq. (20) is defined as

$$\mathcal{I}'_{nl}(q) = \int_0^\infty dr\ r^2 R_{nl}(r) j_l(qr) R_{20}(r). \tag{A1}$$

After performing the integration the electronic radial integral may be expressed as finite sums of Gauss hypergeometric functions, $_2F_1$,

$$\begin{aligned}\mathcal{I}'_{nl}(q) =\ & \frac{1}{q}\frac{2^{l+1/2}}{n^{l+2}(2l+1)!}\left[\frac{(n+l)!}{(n-l-1)!}\right]^{1/2} \text{Re}\left\{\sum_{p=0}^{l}\frac{i^{p-l-1}}{(2q)^p}\frac{(l+p)!}{p!(l-p)!}\right.\\ & \times \sum_{s=0}^{1}\left(-\frac{1}{2}\right)^s (l+1-p+s)!\left(\frac{2n}{2+n-2iqn}\right)^{2+l-p+s}\\ & \left.\times\ _2F_1(l+2+s-p, l-n+1, 2l+2, \frac{4}{2+n-2iqn})\right\}. \end{aligned}\tag{A2}$$

**Appendix B: Atomic radial integrals $\mathcal{J}^a_{nll',20}$ and $\mathcal{J}^b_{nll',20}$**

We recall that the atomic radial integral $\mathcal{J}_{nll'}$ required to calculate the one-photon atomic transition amplitude, $T^{(1)}_{nlm}$, given by Eqs. (29) and (32) is defined as the difference of two atomic radial integrals

$$\mathcal{J}_{nll'}(\tau_2^\mp, \tau_n^\pm, q) = \mathcal{J}^a_{nll',20}(\tau_2^\mp, q) - \mathcal{J}^b_{nll',20}(\tau_n^\pm, q), \tag{B1}$$

with $l' = l \pm 1$ (if $l > 0$) and $l' = 1$ (if $l = 0$) and the dependence of the atomic radial integrals on the energies $\Omega$ is now included in the new parameters $\tau_n^+ = (-2\Omega_n^-)^{-1/2}$ and $\tau_n^- = (-2\Omega_n^+)^{-1/2}$. The two terms in the right-hand side of Eq. (B1) are defined as radial integrals

$$\mathcal{J}^a_{nll',20}(\tau, q) = \int_0^\infty dr\ r^2 R_{nl}(r) j_{l'}(qr) \mathcal{B}_{201}(\tau; r), \tag{B2}$$

and

$$\mathcal{J}^b_{nll',20}(\tau, q) = \int_0^\infty dr\ r^2 R_{20}(r) j_{l'}(qr) \mathcal{B}_{nll'}(\tau; r), \tag{B3}$$

where the radial functions $\mathcal{B}_{nll'}(\tau;r)$ are defined in Ref. [33].

First, we focus on the $\mathcal{J}^a_{nll',20}$ atomic radial integral and after performing some algebra using the developing of the spherical Bessel function in Eq. (B2), the integral is analytically expressed in terms of finite sums of Appell hypergeometric, $F_1$, [45] of two variables

$$\begin{aligned}
\mathcal{J}^a_{nll',20}(\tau,q) &= \frac{\tau}{q} \frac{2^{2\tau+l+9/2}}{n^{2+l}} \frac{1}{(2l+1)!} \left[\frac{(n+l)!}{(n-l-1)!}\right]^{1/2} \\
&\times \text{Re}\left\{ \sum_{p=0}^{l'} \frac{i^{p-l'-1}}{(2q)^p} \frac{(l'+p)!}{p!(l'-p)!} \sum_{s=0}^{n-l-1} \frac{(l+1-n)_s}{(2l+2)_s s!} \left(\frac{2}{n}\right)^s \right. \\
&\times \sum_{\nu=0}^{1} \sum_{\mu=0}^{1} \frac{(4)^\nu}{a} \frac{(-1)_{\mu+\nu}}{(4)_\nu \mu! \nu!} \left(-\frac{1}{4}\right)^\mu \frac{(2-\tau)^{1-\mu-\nu}}{(2+\tau)^{3+\tau-2\mu}} \\
&\times (2+l-p+s+\nu)! \left(\frac{n\tau}{n+\tau-iqn\tau}\right)^{3+l-p+s+\nu} \\
&\left. \times F_1(a, -2-\tau+\mu, 3+l-p+s+\nu, a+1, x_2, y_2) \right\},
\end{aligned} \quad (B4)$$

The expression $(n)_s$, with $n$ and $s$ integers, denotes the Pochhammer's symbol and $a = 2 - \tau + \mu + \nu$. The variables $x_2$ and $y_2$ of the Appell hypergeometric function are

$$x_2 = \frac{2-\tau}{4}, \quad y_2 = \frac{n}{2} \frac{2-\tau}{n+\tau-iqn\tau}. \quad (B5)$$

We should mention that $\mathcal{J}^a_{nll',20}$ presents poles with respect to $\tau$ in Eq. (B4), which arise due to the cancellation of the $a$ denominator and from the poles of the Appell hypergeometric functions for $\tau = n'$, where $n'$ is an integer. The origin of these poles resides in the poles of the Coulomb Green's functions used for the calculation of $\mathbf{w}_{nlm}$ vectors [33]. In particular, $\mathcal{J}^a_{nll',20}(\tau_2^\pm, q)$ presents poles at $\tau_2^- = n'$ with $n' > 2$ and $\tau_2^+ = n'$ with $n' < 2$.

Similarly, the second atomic radial integral $\mathcal{J}^b_{nll',20}$ defined in Eq. (B3) is analytically expressed in terms of finite sums of Appell hypergeometric functions, $F_1$, as

$$\begin{aligned}
\mathcal{J}^b_{nll',20}(\tau,q) &= \frac{\tau}{q} \frac{2^{2l'+\tau+1/2} n^{\tau-1}}{(2l'+1)!} \left[\frac{(n+l)!}{(n-l-1)!}\right]^{1/2} \\
&\times \text{Re}\left\{ \sum_{p=0}^{l'} \frac{i^{p-l'-1}}{(2q)^p} \frac{(l'+p)!}{p!(l'-p)!} \sum_{k=-1,1} \sum_{\nu=0}^{n-l'-1-k} \sum_{\mu=0}^{n-l'-1-k-\nu} d_{n,l}^{l',-k} \right. \\
&\times \frac{(4)^\nu}{b} \left(-\frac{1}{2n}\right)^\mu \frac{(l'+1+k-n)_{\mu+\nu}}{(2l'+2)_\nu \mu! \nu!} \frac{(n+\tau)^{k+2\mu-n-\tau}}{(n-\tau)^{k+\mu+\nu-n+l'+1}} \\
&\times \sum_{s=0}^{1} \left(-\frac{1}{2}\right)^s (l'+1-p+s+\nu)! \left(\frac{2\tau}{2+\tau-2iq\tau}\right)^{2+l'-p+s+\nu} \\
&\left. \times F_1(b, -n-\tau+1+k+\mu, 2+l'-p+s+\nu, b+1, x_n, y_n) \right\}, \quad (B6)
\end{aligned}$$

where the following notations $d_{n,l}^{l+1,1} = (n+l+1)(n+l+2)$, $d_{n,l}^{l-1,1} = 1$, and $d_{n,l}^{l',-1} = -d_{-n,l}^{l',1}$, are introduced in [33], $b = l' + 1 - \tau + \mu + \nu$, and the variables $x_n$ and $y_n$ of the Appell hypergeometric function are

$$x_n = \frac{n-\tau}{2n}, \ y_n = \frac{2}{n}\frac{n-\tau}{2+\tau-2iq\tau}. \tag{B7}$$

Similarly to $\mathcal{J}_{nll',20}^a$, the atomic radial integral $\mathcal{J}_{nll',20}^b$ presents poles with respect to $\tau$ in Eq. (B6), that arise due to the cancellation of the $n - \tau$ and $b$ denominators, as well as from the poles of the Appell hypergeometric function $F_1$ for $\tau = n'$, where $n'$ is an integer. In particular, $\mathcal{J}_{nll',20}^b(\tau_n^\pm, q)$ presents poles at $\tau_n^- = n'$ with $n' > n$ and $\tau_n^+ = n'$ with $2 \leq n' < n$.

### Appendix C: Summation formulas for the vector spherical harmonics $\mathbf{V}_{l'lm}$

The well-known summation formulas of the vector spherical harmonics $\mathbf{V}_{l\pm 1lm}$ [34] used in derivation of DCS given by Eq. (34) and quantities $C_{nl}$ [Eq. (35)] and $D_{nl}$ [Eq. (36)] are presented below:

$$\sum_{m=-l}^{l} Y_{lm}^*(\hat{\mathbf{q}})\mathbf{V}_{l-1lm}(\hat{\mathbf{q}}) = \frac{\sqrt{l(2l+1)}}{4\pi}\hat{\mathbf{q}}, \tag{C1}$$

$$\sum_{m=-l}^{l} Y_{lm}^*(\hat{\mathbf{q}})\mathbf{V}_{l+1lm}(\hat{\mathbf{q}}) = -\frac{\sqrt{(l+1)(2l+1)}}{4\pi}\hat{\mathbf{q}}, \tag{C2}$$

$$\sum_{m=-l}^{l} |\boldsymbol{\epsilon} \cdot \mathbf{V}_{l+1lm}(\hat{\mathbf{q}})|^2 = \frac{1}{8\pi}\left[l|\boldsymbol{\epsilon}|^2 + (l+2)|\boldsymbol{\epsilon} \cdot \hat{\mathbf{q}}|^2\right], \tag{C3}$$

$$\sum_{m=-l}^{l} |\boldsymbol{\epsilon} \cdot \mathbf{V}_{l-1lm}(\hat{\mathbf{q}})|^2 = \frac{1}{8\pi}\left[(l+1)|\boldsymbol{\epsilon}|^2 + (l-1)|\boldsymbol{\epsilon} \cdot \hat{\mathbf{q}}|^2\right], \tag{C4}$$

$$\sum_{m=-l}^{l} [\boldsymbol{\epsilon} \cdot \mathbf{V}_{l+1lm}(\hat{\mathbf{q}})]^*[\boldsymbol{\epsilon} \cdot \mathbf{V}_{l-1lm}(\hat{\mathbf{q}})] = \frac{\sqrt{l(l+1)}}{8\pi}\left[|\boldsymbol{\epsilon}|^2 - 3|\boldsymbol{\epsilon} \cdot \hat{\mathbf{q}}|^2\right]. \tag{C5}$$

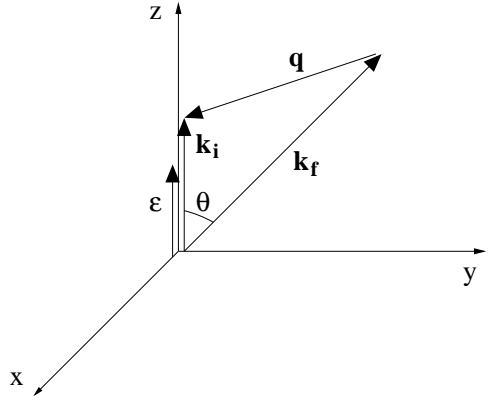

FIG. 1: The specific scattering geometry G1 assumed for the numerical calculations of laser-assisted $e^-$-H($2s$) scattering. $\mathbf{k}_i$ and $\mathbf{k}_f$ are the initial and final momentum vectors of the projectile electron, $\theta$ is the angle between them, and $\mathbf{q}$ is the momentum transfer vector. $\boldsymbol{\epsilon}$ represents the polarization vector of the linearly polarized laser field.

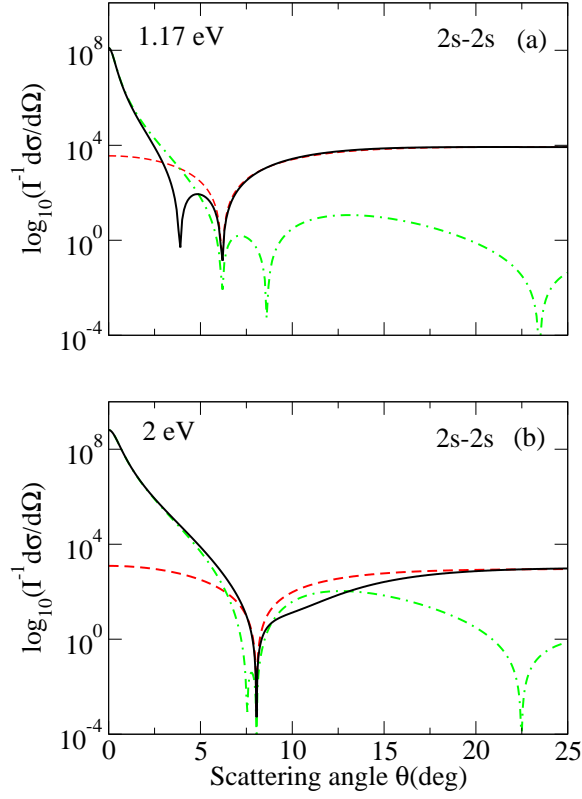

FIG. 2: (Color online) Comparison of the DCS's for the laser-assisted *elastic* scattering of electrons by the H(2s) atoms, with $N = 1$ (one-photon absorption), for incident projectile electron energy $E_{k_i} = 100$ eV, at two photon energies of (a) 1.17 eV and (b) 2 eV, with the laser-atom interaction calculated within the first-order TDPT. The dashed lines represent the Bunkin-Fedorov approximation and the dot-dashed lines represent the atomic contribution to DCS's. The DCS's are normalized by the laser intensity $I$ and the polarization vector of the laser field $\epsilon$ is parallel to the initial momentum vector of the projectile electron $\mathbf{k}_i$.

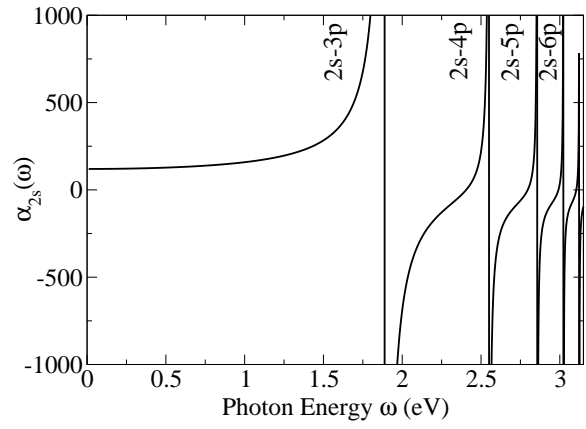

FIG. 3: The dynamic dipole polarizability of the 2s state, $\alpha_{2s}$, calculated from Eq. (42) as a function of the photon energy.

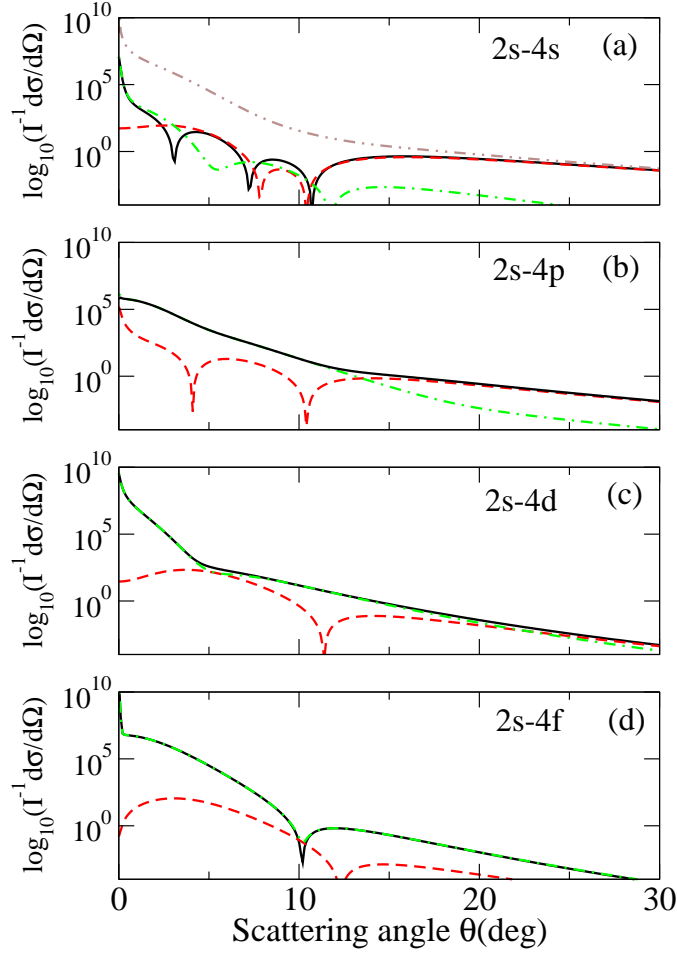

FIG. 4: (Color online) Differential cross sections for laser-assisted $e^-+\text{H}(2s)\to e^-+\text{H}(4l)$ *inelastic* scattering process for $N=1$ as a function of the scattering angle $\theta$, at the projectile electron energy of $E_{k_i} = 500$ eV and photon energy of 1.17 eV with the excitation of the (a) $4s$, (b) $4p$, (c) $4d$, and (d) $4f$ subshells. The dashed lines represent the DCS's given by Eq. (40) where the atomic dressing is neglected and the dot-dashed lines represent the atomic contribution to DCS's. The dot-dot-dashed line in panel (a) represents the total DCS for transition to any $n=4$ subshell. The DCS's are normalized by the laser intensity $I$ and the polarization vector of the laser field $\boldsymbol{\epsilon}$ is parallel to $\mathbf{k}_i$.

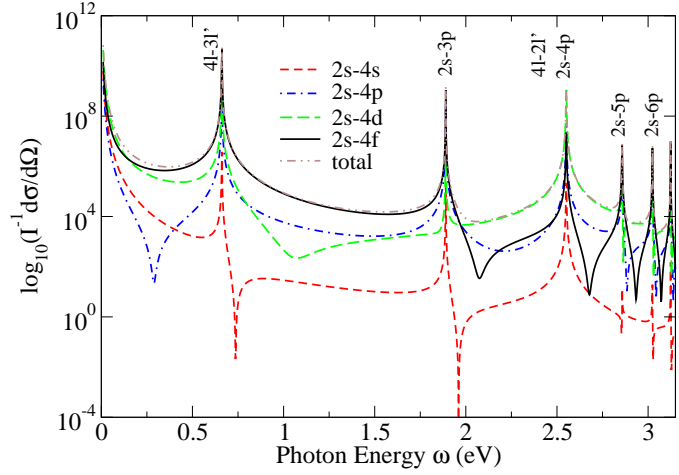

FIG. 5: (Color online) Differential cross sections for laser-assisted $e^-+$H$(2s)\to e^-+$H$(4l)$ *inelastic* scattering process for $N=1$ with the excitation of the $4l$ subshells ($l=0,1,2,$ and 3) as a function of the photon energy $\omega$, at the projectile electron energy of $E_{k_i}=500$ eV and the scattering angle of $5°$, for the $2s\to 4s$ (short dashed line), $2s\to 4p$ (dot-dashed line), $2s\to 4d$ (long dashed line), and $2s\to 4f$ (solid line) excitations of hydrogen. The dot-dot-dashed line represents the total DCS for transition to any $n=4$ subshell. The DCS's are normalized by the laser intensity $I$ and the polarization vector of the laser field $\epsilon$ is parallel to $\mathbf{k}_i$.

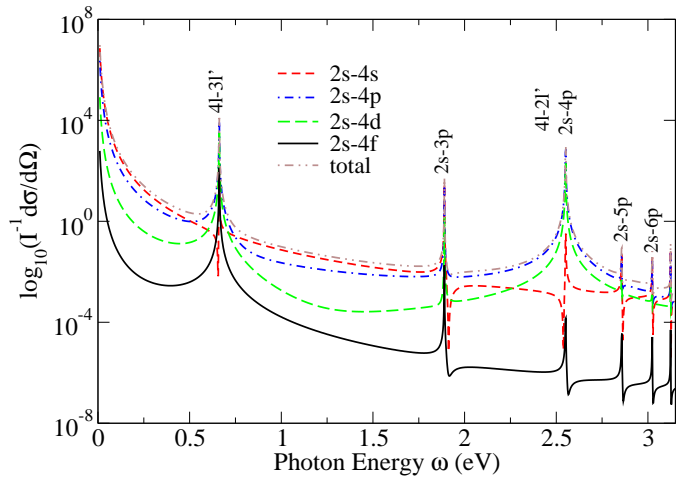

FIG. 6: (Color online) Similar results as those in Fig. 5 but the scattering angle is $30°$.

28